 \documentclass[twocolumn,PRL,aps,psfig,showpacs,preprintnumbers,superscriptaddress]{revtex4}
\usepackage{graphicx}
\usepackage{epstopdf}
\usepackage{float}
\usepackage{color}
\usepackage{amsmath}

\begin{document}

\title{Suppression of ferromagnetic spin fluctuations in the filled skutterudite superconductor SrOs$_4$As$_{12}$ revealed by $^{75}$As NMR-NQR measurements}
\author{Q.-P. Ding}
\affiliation{Ames Laboratory, U.S. DOE, and Department of Physics and Astronomy, Iowa State University, Ames, Iowa 50011, USA}
\author{K. Nishine}
\affiliation{Muroran Institute of Technology, Muroran, Hokkaido 050-8585, Japan}
\author{Y. Kawamura}
\affiliation{Muroran Institute of Technology, Muroran, Hokkaido 050-8585, Japan}
\author{J. Hayashi}
\affiliation{Muroran Institute of Technology, Muroran, Hokkaido 050-8585, Japan}
\author{C. Sekine}
\affiliation{Muroran Institute of Technology, Muroran, Hokkaido 050-8585, Japan}
\author{Y. Furukawa}
\affiliation{Ames Laboratory, U.S. DOE, and Department of Physics and Astronomy, Iowa State University, Ames, Iowa 50011, USA}

\date{\today}

\begin{abstract} 

    Motivated by the recent observation of ferromagnetic spin correlations in the filled skutterudite SrFe$_4$As$_{12}$ [Ding {\it et ~al.} {\it Phys.~Rev.} B {\bf 98}, 155149 (2018)], we have carried out $^{75}$As nuclear magnetic resonance (NMR) and nuclear quadrupole resonance (NQR) measurements to investigate the role of magnetic fluctuations in a newly discovered isostructural superconductor SrOs$_4$As$_{12}$ with a superconducting transition temperature of $T_{\rm c}$ $\sim$ 4.8 K.
    Knight shift $K$ determined by the NQR spectrum under a small magnetic field ($\le$ 0.5 T) is nearly independent of temperature, consistent with the temperature dependence of the magnetic susceptibility.
   The nuclear spin-lattice relaxation rate divided by temperature, 1/$T_1T$, is nearly independent of temperature above $\sim$ 50 K and  increases slightly with decreasing temperature below the temperature.  
  The temperature dependence is reasonably explained by a simple model where a flat band structure with  a small ledge near the Fermi energy is assumed.
   By comparing the present NMR data with those in SrFe$_4$As$_{12}$, we found that  the values of $|K|$ and  $1/T_1T$ in SrOs$_4$As$_{12}$ are smaller than those in  SrFe$_4$As$_{12}$, indicating no obvious ferromagnetic spin correlations in SrOs$_4$As$_{12}$.
    From the temperature dependence of 1/$T_1$ in the superconducting state, an $s$-wave superconductivity is realized.

\end{abstract}

\maketitle

 \section{Introduction} 
 
     Magnetic fluctuation is one of the key parameters to characterize the physical properties of strongly correlated electron systems. 
     Antiferromagnetic spin fluctuations are considered to play an important role in many unconventional superconductors  such as high $T_{\rm c}$ cuprates \cite{HTS}, iron-based superconductors \cite{Johnston2010,Dai2015,Paul2018,DingQCP}, organic superconductors \cite{Organic}, and also heavy fermion superconductors \cite{HF}.
     Alternatively, ferromagnetic spin fluctuations are also considered to be important in the mechanism of superconductivity in U-based superconductors \cite{Aoki2012,Aoki2019}, and  in iron-based superconductors \cite{Johnston2010,Nakai2008,PaulPRB,PaulPRL,Jean2017,Li2019}.
     
    The importance of magnetic fluctuations has also been pointed out in the filled skutterudite compound $A$Fe$_4X_{12}$ ($A$ = alkali metal, alkaline earth metal, lanthanide, and actinide; $X$ = P, and Sb) whose magnetic properties largely depend on the number of valence electrons of the $A$ ions \cite{Leithe-Jasper2003,Leithe-Jasper2004,Schnelle2008,Ishida2005,Matsuoka2005,Matsumura2005,Nakai2005,Matsumura20071,Magishi2014,Schnelle2005,K,Gippius2006,Magishi2006}.
      In the case of $A$ = monovalent Na and K ions with $X$ = Sb, a weak ferromagnetism with a Cuire temperature of $T_{\rm C}$ = 85 K has been observed \cite{Leithe-Jasper2003,Leithe-Jasper2004}, while no magnetic order has been reported for the case of divalent alkaline-earth ions such as Ca, Sr, and Ba, where ferromagnetic spin fluctuations are considered to  play an important role \cite{Schnelle2005,Schnelle2008,Matsuoka2005,K}.
      On the other hand, the importance of antiferromagnetic spin fluctuations has been pointed out in the trivalent ion systems of LaFe$_4$Sb$_{12}$  \cite{Schnelle2008,Gippius2006} and in LaFe$_4$P$_{12}$ \cite{Nakai2005}, although ferromagnetic fluctuations are also reported in LaFe$_4$Sb$_{12}$ \cite{Magishi2006}.

     The effects of magnetic fluctuations on the physical properties of filled skutterudite compounds with different $d$ electrons from 3$d$ to 4$d$ and 5$d$ have also been investigated by replacing Fe by Ru or Os \cite{Takabatake2006,Matsuoka2006,Takegahara2008}. 
      From the magnetization, transport  and specific heat measurements,  the 5$d$-electron compounds of $A$Os$_4$Sb$_{12}$ ($A$ = Sr, Ba)  are found to be placed between the Fe 3$d$ compounds with ferromagnetic spin fluctuations and the Ru 4$d$ compounds having no obvious strong electron correlation effects \cite{Takabatake2006,Matsuoka2006}. 
            Although it is  important to systematically study these physical properties of $d$ electron systems by changing $X$ ions such as P and As for deeper understandings of the role of $d$ electrons, not many studies have been carried out  because of the difficulty in preparing those compounds.

      Recently, new filled-skutterudite arsenide compounds  Sr$T_4$As$_{12}$ ($T$ = Fe, Ru, Os) have been synthesized using a high-pressure synthesis technique \cite {Nishine2017}, which provides a new opportunity of systematic studies of the role of $d$ electrons with  3$d$, 4$d$ and 5$d$.
     For the 5$d$ electron system,  SrOs$_4$As$_{12}$ was  found to  be a new superconductor with a transition temperature of $T_{\rm c}$ = 4.8 K  \cite {Nishine2017}.
     For the 3$d$ and 4$d$ electron systems, on the other hand, SrFe$_4$As$_{12}$  and  SrRu$_4$As$_{12}$ do not exhibit superconductivity down to 2 K, although the electrical resistivities show metallic behavior \cite{Nishine2017}. 
    Magnetization and specific heat measurements \cite{Nishine2017} and theoretical studies \cite{Shankar2017} pointed out a ferromagnetic nature of SrFe$_4$As$_{12}$. 
     Quite recently, ferromagnetic spin correlations were actually reported in the 3$d$-compound SrFe$_4$As$_{12}$ \cite{Ding2018}, similar to the case of $A$Fe$_4$Sb$_{12}$ ($A$ = Ba, Sr) \cite{Schnelle2008,Matsuoka2005,Matsumura2005}.
     Since  the 5$d$-compound SrOs$_4$As$_{12}$ exhibits superconductivity in contrast to the non-superconductors $A$Os$_4$Sb$_{12}$ ($A$ = Ba, Sr) \cite{Evers1994}, it is very interesting to investigate how the magnetic fluctuations changes in the newly discovered superconductor SrOs$_4$As$_{12}$.

    Nuclear magnetic resonance  (NMR) and nuclear quadrupole resonance (NQR) measurements are powerful techniques to investigate the magnetic and electronic properties of materials from a microscopic point of view. 
   It is known that the temperature ($T$) dependence of the nuclear spin-lattice relaxation rate (1/$T_1$) reflects the wave vector $q$-summed dynamical susceptibility. 
   On the other hand, NMR spectrum measurements, in particular the Knight shift $K$, give us information on static magnetic susceptibility $\chi$. 
   Thus from the temperature dependence of 1/$T_1T$ and $K$, one can obtain valuable insights into  magnetic fluctuations in materials. 
    Furthermore, 1/$T_1$ measurements in the superconducting state provide important information in understanding the symmetry of Cooper pairs in superconductors.

    In this paper, we report the results of $^{75}$As NMR, NQR and magnetic susceptibility measurements performed to investigate the magnetic and electronic properties of  SrOs$_4$As$_{12}$ from a microscopic point of view.
   Our experimental data  indicate the strong suppressions of ferromagnetic spin correlations in the superconductor  SrOs$_4$As$_{12}$,  in comparison with SrFe$_4$As$_{12}$ which exhibits   electron correlations enhanced  around ferromagnetic wavenumber $q$ = 0 \cite{Ding2018}.
    The temperature dependence of 1/$T_1$ in the superconducting state evidences an $s$-wave superconductivity in SrOs$_4$As$_{12}$.

 \section{Experimental}
 
    Polycrystalline SrOs$_4$As$_{12}$ samples  were prepared at high temperatures and high pressures using a Kawai-type double-stage multianvil high-pressure apparatus \cite {Nishine2017}. 
   The lattice constant of SrOs$_4$As$_{12}$ with a body centered cubic structure  (space group: $Im\overline{3}$) has been determined by x-ray diffraction measurements to be 8.561 \AA~\cite{Nishine2017}.   
  Magnetic susceptibility measurement was performed using a magnetic properties measurement system (MPMS, Quantum Design, Inc.) under a magnetic field of 5 T.
    NMR and NQR measurements of $^{75}$As ($I$ = $\frac{3}{2}$, $\frac{\gamma_{\rm N}}{2\pi}$ = 7.2919 MHz/T, $Q=$ 0.29 barns) nuclei were conducted using a lab-built phase-coherent spin-echo pulse spectrometer.
   The $^{75}$As-NMR spectra were obtained by sweeping the magnetic field $H$ at a fixed frequency $f$ = 37 MHz, while $^{75}$As-NQR spectra were measured in steps of frequency by measuring the intensity of the Hahn spin echo.  
   The $^{75}$As nuclear spin-lattice relaxation rate (1/$T_{\rm 1}$) was measured with a saturation recovery method.
   $1/T_1$ at each temperature was determined by fitting the nuclear magnetization $M$ versus time $t$  using the exponential function $1-M(t)/M(\infty) = e^ {-(3t/T_{1})^{\beta}}$ for $^{75}$As NQR,  where $M(t)$ and $M(\infty)$ are the nuclear magnetization at time $t$ after the saturation and the equilibrium nuclear magnetization at $t$ $\rightarrow$ $\infty$, respectively.  
   $\beta$ $\sim$ 0.8 is nearly independent of temperature in the paramagnetic state, however, the values of $\beta$ show a complicated temperature dependence in the superconducting state below 4.8 K, as will be discussed later.

\section{$^{75}$As NMR and NQR spectra}

\begin{figure}[tb]
\includegraphics[width=\columnwidth]{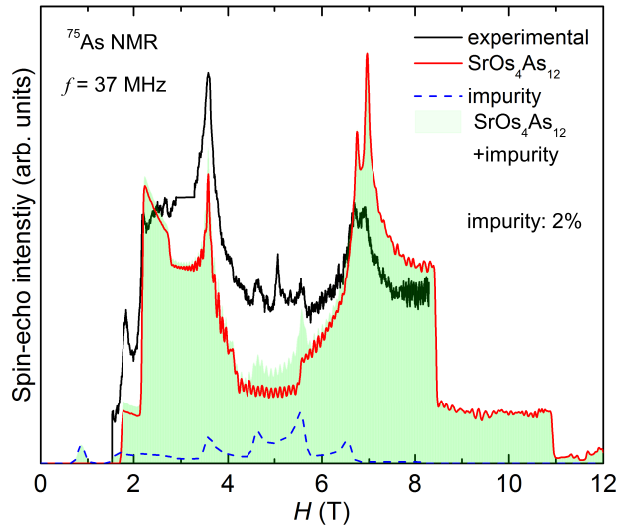} 
\caption{Field-swept $^{75}$As-NMR spectra of SrOs$_4$As$_{12}$ at $f$ = 37 MHz and $T$ = 4.3 K.
The black curve is the observed spectrum and the red curve is the calculated spectrum with $\nu_{\rm Q}$ = 60.1 MHz, $\eta$ = 0.45. 
The blue dotted curve represents the calculated $^{75}$As NMR spectrum ($\nu_{\rm Q}$ = 23.5 MHz, $\eta$ = 0) from the impurity phase \cite{Ding2018}.
The sum of the two calculated spectra is shown by the green area. }
\label{fig:NMR}
\end{figure}   

      Figure\ \ref{fig:NMR} shows the field-swept $^{75}$As-NMR spectrum of SrOs$_4$As$_{12}$ at $T$ = 4.3 K where a complex NMR spectrum is observed. 
     A similar complicated $^{75}$As NMR spectrum has been observed in the  isostructural compound SrFe$_4$As$_{12}$, which is due to a large nuclear quadrupolar interaction and a finite asymmetric parameter $\eta$ of the electric field gradient  (EFG) tensor at the As site \cite{Ding2018}.
    One can calculate NMR spectrum from a nuclear spin Hamiltonian which is  a sum of the nuclear Zeeman  (${\cal H}_{\rm M}$)  and electric quadrupole (${\cal H}_{\rm Q}$) interactions:
\begin{equation}
 {\cal H} ={\cal H}_{\rm M}+{\cal H}_{\rm Q}, 
\label{eq:hamitonian1}
\end{equation}
where 
\begin{equation}
 {\cal H}_{\rm M} =-\gamma\hbar(1+K)H[\frac{1}{2}(I_+ e^{-i\phi}+I_- e^{i\phi}){\rm sin}\theta+I_Z{\rm cos}\theta] 
\label{eq:hamitonian2}
\end{equation}
and 
\begin{equation}
{\cal H}_{\rm Q} = \frac{h \nu_{\rm Q}}{6} [3I_{Z}^{2}-I^2 + \frac{1}{2}\eta(I_+^2 +I_-^2)] ,
\label{eq:hamitonian3}
\end{equation}
 in the coordinate of the principal $X$, $Y$, and $Z$ axes of electric field gradient (EFG).
Here  $H$ is the applied field, $h$ is Planck's constant, $K$ is Knight shift, $\nu_{\rm Q}$ is nuclear quadrupole frequency defined by $\nu_{\rm{Q}}$ =  $eQV_{ZZ}$/2$h$ where $Q$ is the quadrupole moment of the As nucleus, $V_{ZZ}$ is the EFG at the As site,   $\eta$ = $\frac{V_{YY} -V_{XX}}{V_{ZZ}}$ is the asymmetry parameter of the EFG, and $\theta$ and $\phi$ are the polar and azimuthal angles between the direction of the applied field and the $Z$ axis of EFG, respectively. 
     As has been done in our previous paper \cite{Ding2018}, we have  calculated a powder-pattern NMR spectrum by diagonalizing exactly the nuclear spin Hamiltonian without using perturbation theory. 
      The calculated spectrum with the NMR frequency $f$ = 37 MHz, $K$ = 0,  NQR frequency $\nu_{\rm Q}$ = 60.1 MHz and  $\eta$ = 0.45 reasonably reproduces  the characteristic shape of  the observed spectrum, as shown by the red curve in Fig.\ \ref{fig:NMR}.
    However, we notice that, in addition to the calculated spectrum (red curve), there is another contribution ($\sim$~2~$\%$ spectral weight)  of  $^{75}$As NMR spectrum with $\nu_{\rm Q}$ = 23.6 MHz and $\eta$ = 0 to the total NMR spectrum. 
    Similar contribution of the spectrum has been observed in SrFe$_4$As$_{12}$ where the contribution has been  assigned to the impurity phase of  arsenic metal \cite{Ding2018}.


    As for the  principal axis of EFG at the As site, one cannot determine it from NMR spectrum measurements on the powder compound. 
   As described in our previous paper \cite{Ding2018}, Tou {\it et ~al.} \cite{Tou2005, Tou2011} have determined the principal axis of the EFG at the Sb sites in the isostructural PrOs$_4$Sb$_{12}$ compound from NMR measurements using a single crystal, which reports that, although there is one crystallographically equivalent Sb site in the filled skutterudite structure, there are  three different Sb sites with the principal axis  parallel to [100], [010], and [001] of the crystal, respectively, due to the local symmetry of the 24g site of the Sb ions. 
   The same conclusion for the direction of the EFG at the Sb sites in CeOs$_4$Sb$_{12}$ has been reported from $^{121}$Sb NMR using an oriented powder sample \cite{Yogi2009}.  
   Since the crystal structure of the Sb compounds is the same as that of SrOs$_4$As$_{12}$, we consider the directions of EFG at the As sites are the same.


 \begin{figure}[tb]
\includegraphics[width=\columnwidth]{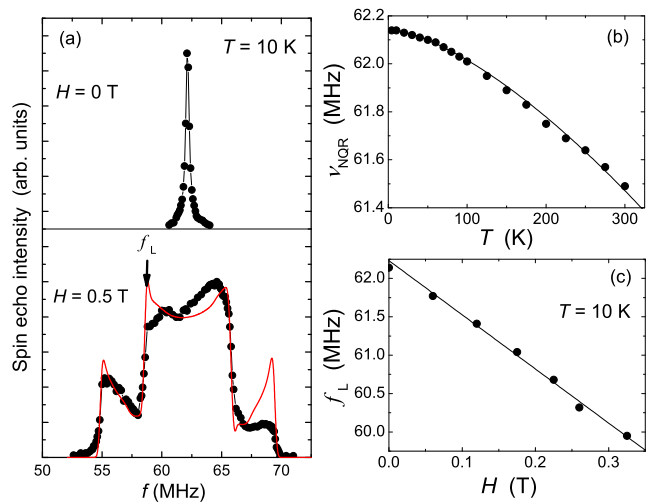} 
\caption{(a) $^{75}$As-NQR spectra measured at $T$ = 10 K under zero magnetic field (upper panel) and 0.5 T (lower panel) in  SrOs$_4$As$_{12}$. 
The red curve at the lower panel is a simulated powder-pattern spectrum with $\nu_{\rm Q}$ = 60.1 MHz, $\eta$ = 0.45 and $H$ = 0.5 T. 
   The  arrow shows the position of the lower-frequency edge position ($f_{\rm L}$) whose position attributes to $\theta$ = $\pi$ (and also $\theta$ = 0). 
(b) $T$ dependence of $^{75}$As-NQR frequency $\nu_{\rm NQR}$ in SrOs$_4$As$_{12}$.
     The black curve is  $\nu_{\rm NQR}(T) = \nu_{\rm NQR}(0)(1-\alpha_{\rm Q} T^{3/2})$  with $\alpha_{\rm Q} = 2.09 \times 10^{-6}$  K$^{-3/2}$ and $\nu_{\rm NQR}(0)$ = 62.14 MHz.
(c) The external magnetic field dependence of $f_{\rm L}$ at $T$ = 10 K. 
}
\label{fig:NQR}
\end{figure}

      In NQR spectrum under zero magnetic field for $I$ = 3/2 where ${\cal H}$ has only ${\cal H}_{\rm Q}$, one can observe  a single transition line at a frequency of  $\nu_{\rm NQR} = \nu_{\rm Q}\sqrt{1+{\eta}^2/3}$.   
      Using the $\nu_{\rm Q}$ = 60.1 MHz and  $\eta$ = 0.45 at $T$ = 10 K estimated from the analysis of the NMR spectrum, one expects the NQR line at $f$ $\sim$ 62.1 MHz which is actually observed as shown at the upper panel in  Fig.~\ \ref{fig:NQR} (a). 
     The peak position slightly shifts to lower frequency with increasing temperature, corresponding to the decrease in $\nu_{\rm NQR}$ on increasing  temperature  as shown in Fig.~\ \ref{fig:NQR} (b).
    Similar temperature dependence of $\nu_{\rm NQR}$ has been observed in SrFe$_4$As$_{12}$ \cite{Ding2018} and also in other filled skutterudite compounds \cite{Matsumura2007,Shimizu2007,Magishi2014,Nowak2009,Nowak2011,Yogi2014} where the temperature dependence at higher temperatures is found to obey an empirical relation $\nu_{\rm NQR}(T) = \nu_{\rm NQR}(0)(1-\alpha_{\rm Q} T^{3/2})$ with a fitting parameter $\alpha_{\rm Q}$.
   This temperature dependence is considered to be due to thermal lattice expansion \cite{alphaQ}.
     As shown by the solid curve in Fig.\ \ref{fig:NQR} (b), the temperature dependence of  $\nu_{\rm NQR}$ in SrOs$_4$As$_{12}$ also follows the relation with $\alpha_{\rm Q} = 2.09 \times 10^{-6}$  K$^{-3/2}$. 
    The value of $\alpha_{\rm Q} = 2.09 \times 10^{-6}$  K$^{-3/2}$ is slightly smaller than 3.21 $\times 10^{-6}$  K$^{-3/2}$ for SrFe$_4$As$_{12}$ \cite{Ding2018}. 
   The linewidth of the NQR spectra (full-width at half maximum, $FWHM$ $\sim$ 0.4 MHz)  is nearly independent of temperature from 4.3 K to 300 K.
   This  indicates that there is no any structural or magnetic phase transition in the normal state of  SrOs$_4$As$_{12}$.


 \section{$^{75}$As Knight shift} 

    Determination of the Knight shift $K$  from the complex NMR spectrum shown in Fig.\ \ref{fig:NMR} is relatively difficult due to the strong quadrupole interaction and relatively large asymmetric parameter $\eta$ value. 
   This is due to the fact that small changes in $\nu_{\rm Q}$ and ${\eta}$ produce a change in $K$, making large ambiguity in determining $K$ from the simulation of NMR spectrum.
    However, as has been reported \cite{Ding2018}, we have succeeded in determining the Knight shift data from NQR spectrum under small magnetic fields lower than 0.5 T in SrFe$_4$As$_{12}$. 
    Here we have applied the same method to SrOs$_4$As$_{12}$ to obtain the Knight shift data. 
As described in detail in Refs. \cite{K,Ding2018}, 
    the NQR resonance frequency [$\nu_{\rm NQR}(H)$] under a small magnetic field can be written by \cite{Dean1954}
\begin{equation}
    \nu_{\rm NQR}(H) = \nu_{\rm NQR}(0) \pm \frac{\gamma_{\rm N}}{2\pi}A(\eta) (1+K)HF(\theta),
\label{eq:NQR_1}
\end{equation}
where  $\nu_{\rm NQR}(0)$ is  $\nu_{\rm NQR}$ at $H$ =  0, $F(\theta)$ = $\frac{{\rm cos}\theta}{2}$$[3-(4{\rm tan}^2\theta+1)^{1/2}]$, and $A(\eta)$ is a factor close to unity  which depends on the value of $\eta$. 
   Under magnetic fields, the random distribution of $\theta$  produces the rectangular shape of the powder-pattern spectrum where $\theta$ = 0 (and also  $\pi$) produce both higher- and lower-frequency edges. 
       By measuring the external magnetic field dependence of the edge position of the NQR spectrum,  one can determine the coefficient of the second term of eq. (\ref{eq:NQR_1}), $A(\eta)$$\frac{\gamma_{\rm N}}{2\pi} (1+K)$, and thus the Knight shift if one knows the value of  $A(\eta)$ since the values of $\frac{\gamma_{\rm N}}{2\pi}$ is known.  
     At the lower panel of Fig. \ \ref{fig:NQR} (a), a typical NQR spectrum observed at $H$ = 0.5 T is shown,  where the rectangular shape of the powder pattern spectrum  is clearly seen. 
    The small peaks (at $\sim$ 55 and 70 MHz) on both sides of the central rectangular spectrum are due to the mixing of states $|$$1/2\rangle$ and $|$$-1/2\rangle$ as a result of zero-order mixing effect \cite{PQR}.
    These features of the observed spectrum are relatively well reproduced by the calculated powder-pattern spectrum  with $\nu_{\rm Q}$ = 60.1 MHz, $\eta$ = 0.45 and $H$ = 0.5 T using the Hamiltonian of eq. (1), as shown by the red curve in Fig. \ \ref{fig:NQR} (a).

    Fig.\ \ref{fig:NQR} (c) shows a typical magnetic field dependence of $f_{\rm L}$  at the lower edge position indicated by the black arrow in  Fig.\ \ref{fig:NQR} (a), exhibiting a clear linear behavior.
   From the slope,  the Knight shift $K$ at 10 K was determined to be -1.2 $\pm$  2.5 $\%$.
   Here we used $A(\eta)$ = 0.9794 as in the case of SrFe$_4$As$_{12}$ \cite{Ding2018}  where the value of $\eta$ = 0.4 is  close to $\sim$ 0.45. 
   Although the error is relatively large, this is much better than the case of NMR spectrum from which we could not determine $K$. 
    It is also noted that we did not include any anisotropy in the Knight shift in the calculated spectrum which reproduces the observed one as shown above. 
   This suggests that, although one expects an anisotropic part in the Knight shift due to the local symmetry of the Os ions (trigonal), 
 the anisotropy is not significant and could not be detected within our experimental uncertainty. 
    Therefore, the Knight shift discussed below is considered as an isotropic part of Knight shift.

\begin{figure}[tb]
\includegraphics[width=\columnwidth]{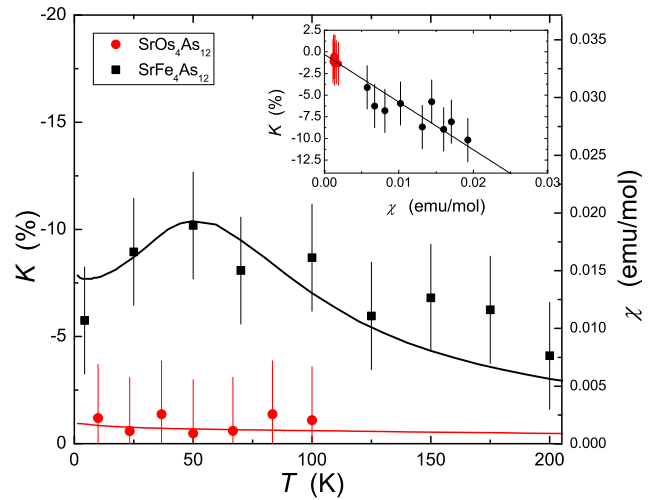} 
\caption{  Temperature dependence of  $^{75}$As Knight shift $K$ in SrOs$_4$As$_{12}$ (red circles) and SrFe$_4$As$_{12}$ (black squares) from Ref. [\onlinecite{Ding2018}]. 
 Temperature dependencies of magnetic susceptibility $\chi$ are also plotted by the red and black curves for  SrOs$_4$As$_{12}$ and SrFe$_4$As$_{12}$ \cite{Nishine2017}, respectively.
   The inset shows $K$ versus corresponding magnetic susceptibility $\chi$.  The black line is a linear fit.
}
\label{fig:K}
\end{figure}

     Following the method described above, we determined $K$ at different temperature. 
   Its temperature dependence is shown in Fig.~\ \ref{fig:K} where the $K$ data in SrFe$_4$As$_{12}$ are also plotted \cite{Ding2018}.
   In contrast to the case of SrFe$_4$As$_{12}$ where the broad minimum around $\sim$ 50 K has been observed (note the sign of the vertical axis of  Fig.~\ \ref{fig:K} for the Knight shifts), $|K|$ values of SrOs$_4$As$_{12}$ are much smaller than those of SrFe$_4$As$_{12}$ and are nearly independent of temperature. 
   These results are consistent with the magnetic susceptibility $\chi$ data whose temperature dependencies are also shown by solid curves in Fig. \ref{fig:K} where the maximum in $\chi$ corresponds to the minimum in $K$ in SrFe$_4$As$_{12}$ due to the negative hyperfine coupling constant \cite{Ding2018}.
   It is also noted that the values of $\chi$ in SrOs$_4$As$_{12}$ is more than one order of magnitude smaller than those in SrFe$_4$As$_{12}$.
   These results clearly evidence that the spin susceptibility of SrOs$_4$As$_{12}$  is much smaller than that of SrFe$_4$As$_{12}$. 
    
    In order to check whether or not the small values of $|K|$ actually correspond to the suppression of spin susceptibility, we have plotted  Knight shifts for both compounds as a function of the corresponding $\chi$ as shown in the inset of Fig. \ref{fig:K}. 
    Since NMR shift consists of temperature dependent spin shift $K_{\rm s}(T)$ and $T$ independent orbital shift $K_{\rm 0}$,  one needs to know $K_{\rm 0}$ to estimate $K_{\rm s}(T)$. 
     $K_{\rm 0}$ can be obtained from the intercept of the so-called $K - \chi$ plot shown in the inset.
    The solid line in the inset is the linear fit for data including both compounds, giving a nearly zero intercept, i.e., nearly zero $K_{\rm 0}$.
     Therefore, the observed $|K|$ can be considered as mainly $K_{\rm s}$, again indicating that the strong suppression of the spin susceptibility in SrOs$_4$As$_{12}$. 
     The slope gives a hyperfine coupling constant $A_{\rm hf}$ = -3.84  $\pm$ 1.60  kOe/$\mu_{\rm B}$ \cite{Hyperfine coupling}.

\begin{figure}[tb]
\includegraphics[width=\columnwidth]{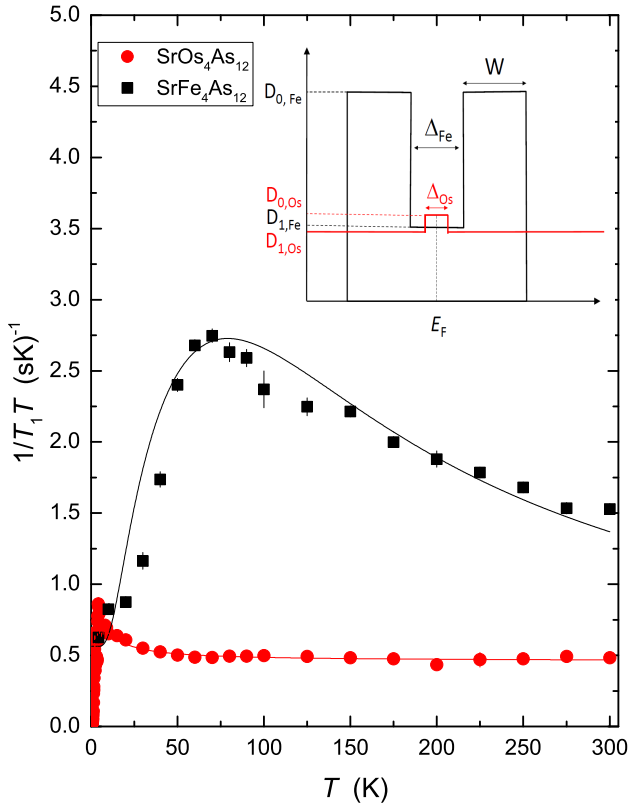} 
\caption{Temperature dependencies of  $^{75}$As 1/$T_1T$ in SrOs$_4$As$_{12}$ (red circles) and SrFe$_4$As$_{12}$ (black squares) from Ref. [\onlinecite{Ding2018}]. 
  The solid lines are the calculated results based on the  band structures near $E_{\rm F}$ shown in the inset with a set of parameters:  $\Delta_{\rm Os}$ = 40 K and ${\cal D}_{0, \rm Os}$/${\cal D}_{1, \rm Os}$ = 1.2, for SrOs$_4$As$_{12}$, and $\Delta_{\rm Fe}$ = 88 K, $W$ = 220 K, and $D_{0, \rm Fe}$/$D_{1, \rm Fe}$ = 2.6 for SrFe$_4$As$_{12}$. 
}
\label{fig:T1}
\end{figure}


\section{$^{75}$As spin-lattice relaxation rate 1/$T_1$} 
\subsection{Normal state}

        Figure \ \ref{fig:T1} shows the temperature dependence of the $^{75}$As  spin-lattice relaxation rate divided by temperature (1/$T_1T$) measured at the peak positions of the NQR spectra under zero magnetic field, together with the data for SrFe$_4$As$_{12}$ reported previously \cite{Ding2018}.
       Contrary to the case in SrFe$_4$As$_{12}$ where the large enhancement of 1/$T_1T$ at low temperatures has been observed, 1/$T_1T$ in SrOs$_4$As$_{12}$ seems follow Korringa law, $T_1T$ = constant,  at high temperature above $\sim$ 50 K though the small increase can be observed at low temperatures in the normal state.

    In the previous paper, to reproduce the temperature dependence of 1/$T_1T$ for SrFe$_4$As$_{12}$,   we have employed a simple model where a concave-shaped band structure shown in the inset of Fig.~\ \ref{fig:T1} is assumed \cite{Ding2018}. 
     In the model, the Fermi energy ($E_{\rm F}$) is assumed to be at the center of the dip, and $\Delta_{\rm Fe}$, $W$, the density of states ${\cal D}_{0, \rm Fe}$ and ${\cal D}_{1, \rm Fe}$ characterize the band structure near $E_{\rm F}$. 
      Using the formula, 
\begin{equation}
\frac{1}{T_1}\sim\int_0^\infty{\cal D}^2(E)f(E)(1-f(E)){\rm{d}}E 
\label{eq:T1}
\end{equation}
where $f(E)$ is the Fermi distribution function, we calculated $1/T_1T$ with a set of parameters of $\Delta_{\rm Fe}$ = 88 K, $W$ = 220 K, and  ${\cal D}_{0, \rm Fe}$/${\cal D}_{1, \rm Fe}$ = 2.63, which reasonably reproduces the experimental data, as shown by the black curve in Fig. \ \ref{fig:T1}.  
 The model is also used to reproduce the temperature dependence of the magnetic susceptibility in SrFe$_4$As$_{12}$ \cite{Ding2018}. It is interesting to note that a similar model (but with no finite density of state at $E_{\rm F}$) has been applied to FeSi for explaining the characteristic temperature dependence of magnetic susceptibility, specific heat and thermal expansion \cite{Mandrus1995}. 

    In the case of SrOs$_4$As$_{12}$, the nearly temperature independent behavior of 1/$T_1T$ indicates a almost flat band near $E_{\rm f}$, contrary to the case of SrFe$_4$As$_{12}$.
    In fact, as shown in by the red solid curve in Fig. \ref{fig:T1},  the observed temperature dependence of 1/$T_1T$ can be reproduced by the band model with a nearly flat structure  near $E_{\rm F}$ with a small ledge having the width $\Delta_{\rm Os}$ = 40 K with ${\cal D}_{0, \rm Os}$/${\cal D}_{1, \rm Os}$ = 1.2. 
    It turns out that the difference in the band structure produces the significantly different behavior in 1/$T_1T$ and could originate from the fact that Fe 3$d$ electrons have more localized nature than Os 5$d$ electrons.


    Now we  discuss magnetic correlations in both systems based on the NMR data.
    As has been discussed in our previous paper, in order to discuss the magnetic correlations,  it is useful to estimate the quantity $T_1TK_{\rm s}^2$ where $K_ {\rm s}$  is the spin part of the Knight shift \cite{Moriya1963,Narath1968,Li2010}. 
 The so-called Korringa ratio 
 ${\cal K}$$(\alpha)\equiv$ $\frac{ \cal S}{T_1TK_{\rm s}^2}$
is unity for uncorrelated metals. 
   Here  ${\cal S}$ =  $\frac{\hbar}{4\pi k_{\rm B}} \left(\frac{\gamma_{\rm e}}{\gamma_{\rm N}}\right)^2$ where $\gamma_{\rm e}$ and $\gamma_{\rm n}$ are the electron and nuclear gyromagnetic ratios, respectively. 
   Since 1/$T_1T$ probes the dynamical spin susceptibility averaged over the Brillouin zone, it can be enhanced either ferromagnetic or antiferromagnetic spin correlations.   On the contrary, $K_{\rm s}$ will be enhanced only for ferromagnetic spin correlations. Therefore, for antiferromagnetic correlated metals,  ${\cal K}$$(\alpha)$  is expected to be greater than unity. On the other hand, one can expect ${\cal K}$$(\alpha)$  much smaller than unity for ferromagnetic spin correlations.
   ${\cal K}$$(\alpha)$ $\sim$ 0.02, much less than unity, was reported in SrFe$_4$As$_{12}$ \cite{Ding2018}, evidencing the ferromagnetic spin correlations.
    On th other hand, the small values of $|K|$ for SrOs$_4$As$_{12}$ increase the  ${\cal K}$$(\alpha)$ values, suggesting much less ferromagnetic spin correlations in SrOs$_4$As$_{12}$.   
   However, the estimate of  ${\cal K}$$(\alpha)$ is rather difficult due to large errors in $K$. In addition, it should be noted that the observed 1/$T_1T$ is the sum of two contributions: the spin and orbital relaxation rates. This also makes difficulty in estimating ${\cal K}$$(\alpha)$.  
    Assuming ${\cal K}$$(\alpha)$  = 1 for uncorrelated metals, $|K_{\rm s}|$ is estimated to be 0.21 $\%$ using the 1/$T_1T$  values above 50 K in the normal state. 
   The $K_{\rm s}$ would be consistent with the observed small values of $|K|$ for SrOs$_4$As$_{12}$ within our large experimental uncertainty. 
   Thus one can conclude that the ferromagnetic spin correlations observed in SrFe$_4$As$_{12}$ are strongly suppressed in SrOs$_4$As$_{12}$  \cite{T1}.

   It is interesting to point out that the 1/$T_1T$ values for both systems are almost comparable at low temperatures below $\sim$ 20 K in the normal state. 
   This indicates that the effective density of states at $E_{\rm F}$, $\cal D(E_{\rm F})$,  is nearly the same for the compounds at low temperatures, as actually illustrated in the inset of Fig. \ref{fig:T1}.
   Since $\cal D(E_{\rm F})$ generally correlates with $T_{\rm c}$ in conventional BCS superconductors, one may expect an appearance of superconductivity  in SrFe$_4$As$_{12}$.
   No superconductivity is, however,  observed in SrFe$_4$As$_{12}$. 
   Therefore,  we speculate that the strong ferromagnetic fluctuations prevent superconductivity.  
   As will be discussed below, the superconductivity in SrOs$_4$As$_{12}$ is revealed to be a spin-singlet $s$-wave state where ferromagnetic fluctuations may compete since it would be favorable to inducing triplet Cooper pairs.
   In other word, the strong suppression of the ferromagnetic spin correlations could  induce the superconductivity in  SrOs$_4$As$_{12}$. 
    Therefore, one may expect the superconductivity in SrFe$_4$As$_{12}$ if the ferromagnetic correlations could be suppressed. 
   This would be possible by applying pressure since the nature of localization of 3$d$ electrons responsible for the ferromagnetic spin correlations may be decreased.
   This interesting project  is currently in progress.

       \subsection{Superconducting state}

\begin{figure}[tb]
\includegraphics[width=\columnwidth]{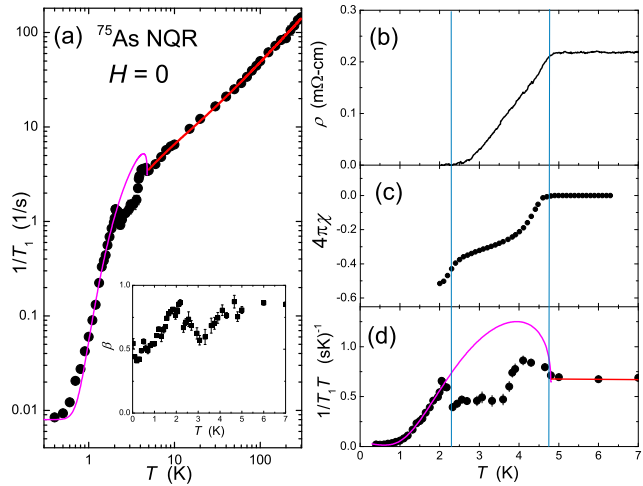} 
\caption{ (a) Temperature dependence of 1/$T_1$ in SrOs$_4$As$_{12}$. 
    The pink curve in the superconducting state is the calculated result based on the BCS theory 
   The red curve in the normal state is the calculated results shown in Fig. \ref{fig:T1}. 
     The inset show the temperature dependence of  $\beta$. 
(b) Temperature dependence of electrical resistivity from Ref. [\onlinecite{Nishine2017}].
(c)  Temperature dependence of volume magnetic susceptibility in the superconducting state estimated from the zero-field cooled magnetic susceptibility data reported in Ref. [\onlinecite{Nishine2017}].
(d)  Temperature dependence of 1/$T_1T$ at low temperature region together with the calculated result (pink curve) based on BCS theory. 
     The vertical blue lines are just guides to the eye.
}
\label{fig:T1T_SC}
\end{figure}

       Finally we show the $T_1$ data in the superconducting  state below $T_{\rm c}$ $\sim$ 4.8~K. 
      As shown in Fig.\ \ref{fig:T1T_SC} (a), 1/$T_1$ shows a tiny, but visible, coherence peak (also known as Hebel-Slichter peak) just below $T_{\rm c}$, and decreases below 4 K, then exhibits a small peak around 2 K following the sudden decrease below $\sim$ 2 K where 1/$T_1$ seems to level off below $\sim$ 0.4 K. 
      Although the temperature dependence of 1/$T_1$ is complicated, it is important to point out that 1/$T_1$ decreases largely by more than two orders of magnitude below $T_{\rm c}$ down to the lowest temperature $\sim$ 0.4 K. 
      This clearly indicates that the superconductivity observed in SrOs$_4$As$_{12}$ is not filamentary or surface, but bulk in nature at least below $\sim$2 K.
      In addition, the observation of the coherence peak just below $T_{\rm c}$ [more clearly seen in Fig. \ref{fig:T1T_SC} (d)] is direct evidence of a conventional $s$-wave BCS superconductor.      
     The observation of the peak also indicates that the relaxation process of the As nuclei in SrOs$_4$As$_{12}$ can be characterized to be magnetic, not electric \cite{Hammond1960, Wada1973}.  
      The unusual and complicated behavior in $1/T_1$ in the temperature range of  2 - 4 K could be due to the distribution of $T_{\rm c}$ in our sample as can be seen in the resistivity and magnetization data \cite{Nishine2017}. 
     As shown in Fig. \ref{fig:T1T_SC} (b), the electrical resistivity starts to decreases below 4.8 K and reaches nearly zero resistivity  at $\sim$ 2.2 K with a relatively broad superconducting-transition width $\Delta$$T_{\rm c}$ = 2.6 K. 
     The distribution of $T_{\rm c}$ can be also inferred from the two step behavior in the volume magnetic susceptibility in the superconducting state estimated from the zero-field cooled magnetic susceptibility measurements as shown in  Fig. \ref{fig:T1T_SC} (c). 
     It is noted that the superconducting volume starts to increase below $\sim$ 4.8 K on cooling and reaches $\sim$ 45 $\%$ around 2.2 K where the zero resistivity is observed. 
     The distribution of $T_{\rm c}$ would make problems in estimating $T_1$ since the observed spectrum may compose of at least two components of $^{75}$As NQR signals from the superconducting and the normal states. 
    In addition, the ratio of the signals from the superconducting and normal states could be changed with temperature due to Meissner effect, which will produce complicated nuclear relaxation curves.
    In fact, as shown in the inset of Fig. \ref{fig:T1T_SC} (a), the temperature dependence of $\beta$ exhibits a complicated behavior with a local minimum around 3.5 K, showing the large distribution of 1/$T_1$. 
   Thus, since we consider that the temperature dependence of 1/$T_1$ in $T$ = $2.2-4$ K is not intrinsic but extrinsic and more artificially, we will not discuss it in this paper.
   It is noted that, although similar complicated behavior of 1/$T_1$ has been observed in SrtPtAs \cite{Bruckner2014}, the later $T_1$ measurements with high quality samples do not show the complicated behavior \cite{Matano2014}, suggesting that the quality of samples strongly affects the temperature dependence of 1/$T_1$.

   In order to test whether the observed temperature dependence of 1/$T_1$ can be explained by a $s$-wave conventional superconductor model, we have calculated 1/$T_1$ using BCS theory.
     Here, the relaxation rate 1/$T_{\rm 1s}$  in the superconducting state normalized by $1/T_{\rm 1n}$ in the normal state is expressed as \cite{Hebel1959}
 \begin{eqnarray}
\frac{T_{\rm 1n}}{T_{\rm 1s}} \propto \int_{0}^{\infty}[{N_{\rm s}(E)}^2+{M_{\rm s}(E)}^2]f(E)[1-f(E)]dE
 \end{eqnarray}
where $M_{\rm s}(E)$ = $N_0\Delta$/$\sqrt{E^2-{\Delta}^2}$ is the anomalous density of states (DOS) due to the coherence factor,  $N_{\rm s}(E)$ = $N_{\rm 0}E$/$\sqrt{E^2-{\Delta}^2}$ is the DOS in the superconducting state, $\Delta$ is the energy gap, $N_0$ is the DOS in the normal state, and $f(E)$ is the Fermi distribution function. 
   We convolute  $M_{\rm s}(E)$ and $N_{\rm s}(E)$ with a broadening function assuming a triangle shape with a width 2$\delta$ and a height 1/$\delta$ \cite{Ding2016}. 
   The  pink curve in Fig. \ref{fig:T1T_SC} (a)  [also shown in Fig. \ref{fig:T1T_SC} (d)] is the calculated result with $\Delta(0)$ = 6 K and  $r=\Delta(0)/\delta=5$.
   Here we added a constant term of 1/$T_1$ = 0.008 (1/s) into the calculated result to reproduce the temperature independent behavior of 1/$T_1$ observed below $\sim$ 0.4 K.  
  The origin of the temperature independent behavior of 1/$T_1$ is not clear at present but it may originate from some sort of impurity effects.  
   Although the calculated results does not reproduce the height of the observed coherence peak, it seems to capture the observed temperature dependence of 1/$T_1$ (here again, except for the intermediate temperature range of $2.2-4$ K).
    The value of 2$\Delta(0)/k_{\rm B}T_{\rm c}$ is estimated to be 2.5 using the on-set $T_{\rm c}$ of 4.8 K. 
   This value is slightly smaller than that of the BCS weak-coupling limit 2$\Delta(0)/k_{\rm B}T_{\rm c}$ = 3.53. 
   It should be noted that the isostructural compound CaOs$_4$P$_{12}$, has been suggested to be a BCS-type superconductor with $T_{\rm c}$ = 2.5 K \cite{Kawamura2018}. 
   In order to discuss the superconducting properties of SrOs$_4$As$_{12}$  in detail, one needs good quality samples with much smaller  distribution of $T_{\rm c}$ and the detailed study of the superconducting properties of SrOs$_4$As$_{12}$ is a future project. 
   However, it should be noted that our NMR measurements clearly indicate the $s$-wave BCS superconducting nature in SrOs$_4$As$_{12}$, even though the present samples show the relatively broad superconducting transition. 
 

   \section{Summary}
      In summary,  we have carried out $^{75}$As NMR and NQR measurements on the superconductor SrOs$_4$As$_{12}$ and compared the results with those in  non-superconducting metallic compound SrFe$_4$As$_{12}$ in which ferromagnetic spin correlations play an important role. 
       Knight shift $K$ determined by the NQR spectrum under a small magnetic field ($\le$ 0.5 T) is nearly independent of temperature, which is
consistent with the temperature dependence of the magnetic susceptibility.
    The values of $|K|$ in SrOs$_4$As$_{12}$ are much smaller than those in SrFe$_4$As$_{12}$, indicating the spin susceptibility in SrFe$_4$As$_{12}$ is strongly suppressed  in SrOs$_4$As$_{12}$.
    Similar strong suppression in SrOs$_4$As$_{12}$ can also be observed in 1/$T_1T$ data whose temperature dependence is explained by a simple model where we assume a flat band structure with  a small ledge near the Fermi energy.
    The large suppression in the $|K|$ and  $1/T_1T$ in SrOs$_4$As$_{12}$ compared with those in SrFe$_4$As$_{12}$ indicates no obvious ferromagnetic spin correlations in SrOs$_4$As$_{12}$.
   Furthermore, the temperature dependence of 1/$T_1$ in the superconducting state evidences a conventional $s$-wave superconductivity in SrOs$_4$As$_{12}$.
    These results suggest that the ferromagnetic spin correlations compete the appearance of superconductivity in the Sr-based filled skutterudite arsenides, which may be consistent with  the conventional spin-singlet $s$-wave superconducting state revealed by 1/$T_1$ measurements.

   \section{Acknowledgments} 
The research was supported by the U.S. Department of Energy (DOE), Office of Basic Energy Sciences, Division of Materials Sciences and Engineering. Ames Laboratory is operated for the U.S. DOE by Iowa State University under Contract No.~DE-AC02-07CH11358.
    Part of this work was supported by JSPS KAKENHI Grant Number 23340092.


\begin{thebibliography}{10}
\bibitem {T1}
     In this analysis, we have assumed that 1/$T_1$ is isotropic. 
     In order to check the anisotropy in 1/$T_1$, we have measured the position dependence of 1/$T_1$ in the As NQR spectrum under a magnetic field of 0.5 T. 
      We do not observe significant change in 1/$T_1$  at different positions (i.e., different $\theta$)  of the spectrum within our experimental uncertainty, indicating that $T_1$ is nearly isotropic.
\end{thebibliography}
\end{document}